\PassOptionsToPackage{hyphens}{url}
\documentclass[format=sigconf,letterpaper,authorversion]{acmart}

\usepackage[hyphens]{url}
\usepackage{hyperref}
\usepackage{lipsum}

\usepackage{booktabs} %

\usepackage{etoolbox}
\usepackage{acro}
\usepackage[american]{babel}

\hypersetup{breaklinks=true}
\usepackage{epsfig}
\usepackage{caption}
\usepackage{subcaption}
\usepackage{ulem}

\usepackage{colortbl}
\usepackage{units}

\usepackage[inline]{enumitem}
\usepackage{balance}
\usepackage{nth}
\usepackage{dblfloatfix}
\usepackage{listings}
\usepackage{tikz}
\usetikzlibrary{automata,backgrounds,fit,positioning,shapes,calc,arrows.meta}

\hypersetup{pdfstartview=FitH,pdfpagelayout=SinglePage}

\makeatletter %

\newcommand{\ie}{i.e.,}
\newcommand{\eg}{e.g.,}
\newcommand{\etal}{et~al\@ifnextchar.{}{.\@}}
\newcommand{\etc}{etc\@ifnextchar.{}{.\@}}

\newcommand{\sref}[1]{Section~\ref{#1}} 
\newcommand{\fig}[1]{Figure~\ref{#1}} 

\newcommand{\afblock}[1]{\noindent{\textbf{#1}}}
\newcommand{\takeaway}[1]{\noindent{\textbf{Takeaway.}} \textit{#1}}

\acmYear{2018}\copyrightyear{2018}
\setcopyright{acmcopyright}
\acmConference[IMC '18]{IMC '18: Internet Measurement Conference}{October 31--November 2, 2018}{Boston, MA, USA}
\acmBooktitle{IMC '18: Internet Measurement Conference, October 31--November 2, 2018, Boston, MA, USA}
\acmPrice{15.00}
\acmDOI{10.1145/3278532.3278539}
\acmISBN{978-1-4503-5619-0/18/10}

\AtEndDocument{\balance}

\def\MoneroMedDiff{{55.4G}}

\def\MoneroMedHashRate{{462M}}

\def\CoinhiveMedBlocksPerDay{{8.5}}

\def\CoinhiveAvgBlocksPerDay{{9.0}}

\def\CoinhiveMedBlocksPerDayShare{{1.18\%}}

\def\CoinhiveMedHashRate{{5.5M}}

\def\CoinhiveAvgUsersLowPerf{{292K}}

\def\CoinhiveAvgUsersHighPerf{{58K}}

\def\CoinhiveXMRRewardPerMonth{{1,271}}

\def\CategoryOneName{{Tech. \& Telecomm.}}

\def\CategoryOneOcc{{1,522}}

\def\CategoryTwoName{{Gaming}}

\def\CategoryTwoOcc{{737}}

\def\CategoryThreeName{{Dynamic Site}}

\def\CategoryThreeOcc{{727}}

\def\CategoryFourName{{Business}}

\def\CategoryFourOcc{{578}}

\def\CategoryFiveName{{Pornography}}

\def\CategoryFiveOcc{{577}}

\def\CategorySixName{{Shopping}}

\def\CategorySixOcc{{572}}

\def\CategorySevenName{{Finance and Investing}}

\def\CategorySevenOcc{{502}}

\def\CategoryEightName{{Ent. \& Music}}

\def\CategoryEightOcc{{313}}

\def\CategoryNineName{{Educational Site}}

\def\CategoryNineOcc{{305}}

\def\CategoryTenName{{Hosting}}

\def\CategoryTenOcc{{298}}

\newcommand{\avgblocksperday}[1]{%
\IfStrEqCase{#1}{%
{july}{9.1}
{june}{9.7}
{may}{8.8}
}%
[nada]%
}
\newcommand{\medblocksperday}[1]{%
\IfStrEqCase{#1}{%
{july}{9.0}
{june}{10.0}
{may}{9.0}
}%
[nada]%
}
\newcommand{\avgblocksshare}[1]{%
\IfStrEqCase{#1}{%
{july}{0.01\%}
{june}{0.01\%}
{may}{0.01\%}
}%
[nada]%
}
\newcommand{\medblocksshare}[1]{%
\IfStrEqCase{#1}{%
{july}{0.01\%}
{june}{0.01\%}
{may}{0.01\%}
}%
[nada]%
}
\newcommand{\avghashrate}[1]{%
\IfStrEqCase{#1}{%
{july}{5.9}
{june}{5.4}
{may}{5.5}
}%
[nada]%
}
\newcommand{\medianhashrate}[1]{%
\IfStrEqCase{#1}{%
{july}{5.8}
{june}{5.5}
{may}{5.5}
}%
[nada]%
}
\newcommand{\xmrminded}[1]{%
\IfStrEqCase{#1}{%
{july}{1215}
{june}{1293}
{may}{1231}
}%
[nada]%
}
\begin{document}
\normalem

\title{Digging into Browser-based Crypto Mining}

\renewcommand{\shortauthors}{}

\author{Jan R\"uth}
\author{Torsten Zimmermann}
\author{Konrad Wolsing}
\author{Oliver Hohlfeld}
\affiliation{%
\institution{Communication and Distributed Systems, RWTH Aachen University, Germany}
}
\email{{lastname}@comsys.rwth-aachen.de}

\renewcommand{\shortauthors}{R\"uth et al.}

\begin{abstract}
Mining is the foundation of blockchain-based cryptocurrencies such as Bitcoin rewarding the miner for finding blocks for new transactions.
The Monero currency enables mining with standard hardware in contrast to special hardware (ASICs) as often used in Bitcoin, paving the way for in-browser mining as a new revenue model for website operators.
In this work, we study the prevalence of this new phenomenon.
We identify and classify mining websites in 138M domains and present a new fingerprinting method which finds up to a factor of 5.7 more miners than publicly available block lists.
Our work identifies and dissects Coinhive as the major browser-mining stakeholder.
Further, we present a new method to associate mined blocks in the Monero blockchain to mining pools and uncover that Coinhive currently contributes \CoinhiveMedBlocksPerDayShare{} of mined blocks having turned over  \xmrminded{june} Moneros in June 2018.
\end{abstract}



\maketitle
\section{Introduction}
\label{sec:intro}
The web economy has traditionally used advertisements as means to monetize services that are offered free of charge.
This business model relies on the implicit agreement between content providers and users where viewing ads is the price for the ``free'' content.
This traditional approach has very recently been complemented by a new monetizing model in which the computational resources of website visitors are used to mine cryptocurrencies to generate revenue for the website operators (browser-based mining).

Mining is the method of producing new blocks in blockchain systems, most prominently cryptocurrencies such as Bitcoin. 
It requires miners to solve a computationally expensive puzzle to cryptographically link a new block to the previous block in the blockchain.
The difficulty to solve this puzzle depends on the combined computing power of all users---depending on the difficulty, an individual requires powerful machines to increase the probability of mining a block (\eg{} GPUs, FPGAs, or even ASICs).
To provide an incentive for contributing computational power, miners are awarded currency for every mined block.
This monetary reward has rendered crypto mining a business---browser-based mining extends this business to monetize the web.

Not all cryptocurrencies are equally suited for browser-based mining.
The hardware imbalance and the consequential high difficulty to mine Bitcoin renders its in-browser mining inefficient and motivates the use of, \eg{} Monero as an alternative currency that can be efficiently mined on CPUs and thus browsers.
Given its design, Monero has been adopted by websites (\eg{} The Piratebay or a video streaming service with subsequent media exposure~\cite{guardianPirateBay,bbcMiningHacking}) and even among botnets to mine on millions of compromised hosts~\cite{moneroBotnet}.
To ease browser mining, APIs~\cite{coinhive,cryptoloot} exist, \eg{} for in-game financing~\cite{BTCManagerCoinhiveInGameFinancing}, link forwarding~\cite{CoinhiveRedirectYoutubeAstley}, captchas, during video streaming~\cite{guardingStreamingMining} or even as an entry fee for parties~\cite{cryptoRave}.
Our work identifies Coinhive~\cite{coinhive} as a widely used service which provides a framework for embedding a Monero miner into a website.
While these frameworks enable mining without the users' knowledge (cryptojacking), other services (Authedmine) actively ask users for their consent to do so as an alternative to displaying ads.
Besides media reports, little is known about the ubiquity and use of browser-based mining.

Given these new possibilities, we provide a first in-depth study of the {\em prevalence} and {\em economics} of browser-based mining as a new web business model.
We base this perspective on crawls of 137M .com/.net/.org domains and the Alexa Top 1M list to first identify sites using browser-based mining enabling to create a new fingerprinting method to identify mining code.
Second, we dissect the short link service of the largest web-mining stakeholder Coinhive and screen their market power and profits.
Our contributions are:
\vspace{-0.55em}
\begin{itemize}[noitemsep,topsep=5pt,leftmargin=9pt]
	\item We investigate the prevalence of browser-mining in the three largest TLDs and the Alexa Top 1M, \ie{} at over 138M domains.
	\item We present a new Wasm-based fingerprinting method showing the inadequate capabilities of block lists to detect mining.
	\item Moreover, we identify the largest browser-based mining provider Coinhive and dissect their link-forwarding service.
	\item We present a novel methodology enabling us to associate blocks in a privacy-preserving blockchain to a mining pool.
	\item By applying our methodology, we screen Coinhive and show that they contribute \CoinhiveMedBlocksPerDayShare{} of the blocks in the Monero blockchain mining Moneros worth 150,000 USD per month (as of writing).
\end{itemize}
\vspace{-0.55em}

\afblock{Structure.}
\sref{sec:background} establishes the basics of mining. 
\sref{sec:world-wide} measures the prevalence of browser-mining.
\sref{sec:coinhive} studies the practices, userbase, and economics of Coinhive.
\sref{sec:rw} discusses related work and \sref{sec:conclusion} concludes the paper.

\section{Browser-based Mining 101}
\label{sec:background}

\begin{figure}
\centering
\includegraphics[]{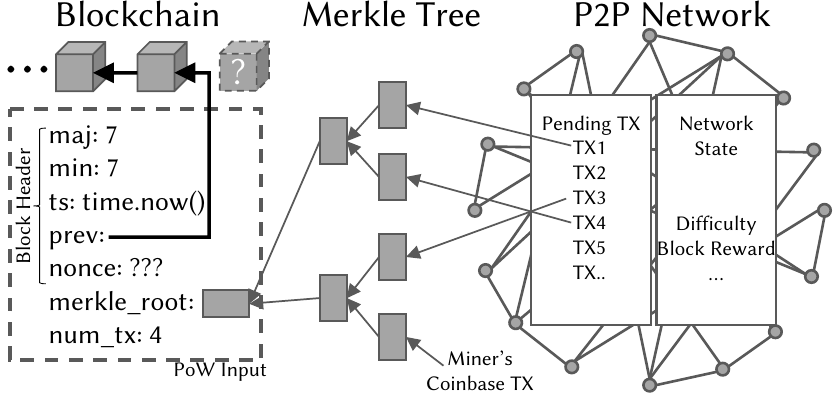}
\vspace{-2em}
\caption{Monero blockchain and PoW mining input.}
\label{fig:monero_overview}
\vspace{-1em}
\end{figure}

Blockchain-based cryptocurrencies build on the principle of embedding financial transactions in a public, tamper-proof series of blocks.
To evolve the system, new blocks must constantly be appended to store pending transactions; their generation is called {\em mining}.
Miners solve a crypto puzzle as a proof of work (PoW) whose {\em difficulty} is dynamically adjusted to produce new blocks at a constant {\em block rate} guaranteeing predictability and tamper resistance.
Consequently, when more miners compete for finding blocks, the difficulty rises such that the block rate is met.
When the PoW meets the difficulty, it links the newly mined block (containing new transactions) to the previous one rewarding the miner with \emph{currency} in exchange for the contributed computing power.

The recent hype around cryptocurrencies has led to substantial increases in difficulty resulting in the need for faster hardware to mine blocks profitably w.r.t. the energy costs.
To increase the chance of earning currency, miners seek to increase their available computational power.
This quest for speed is currently served by GPUs, FPGAs, or even specialized ASICs.
One can host substantial amounts of mining hardware in dedicated data centers.
Another way is to bundle the computing power of multiple miners in {\em mining pools} that share the earned revenue for the newly mined block.

\afblock{Browser-based Mining.} 
Utilizing the computation power of website visitors provides yet another mean of increasing the mining power.
By embedding mining code into websites, a miner can make use of the visitor's CPU resources during the visit.
The website operator thereby saves energy costs and mining hardware investments.
Thus, web-based mining is an alternative revenue generating model to monetize websites and services.
However, hidden mining or without user consent (i.e., cryptojacking) poses a significant challenge and it is a known attack vector (\sref{sec:rw}).
While browser miners for Bitcoin exist (\eg{} jsMiner from 2011~\cite{jsminer}), the performance imbalance between CPUs, GPUs, and ASICs poses an insurmountable challenge for Bitcoin browser mining.
Consequently, browser-based mining requires cryptocurrencies with PoW functions that are only efficiently computable on CPUs.

\afblock{Monero.}
Launched in 2014, Monero~\cite{monero_website} (see \fig{fig:monero_overview}) is a privacy-preserving cryptocurrency whose PoW is designed to be ASIC resistant (memory intensive and periodically redesigned) enabling CPUs and thus browser-based mining.
Specifically, it uses the Cryptonight hash function~\cite{Cryptonight} in its PoW to mine a new block with an average block rate of two minutes.
\fig{fig:monero_overview} shows the PoW inputs; in Monero, a miner constructs a Merkle tree of the transactions that are to be included in the new block, requiring at least the \emph{Coinbase} transaction paying the block reward to the miner.
A node in this tree is the hash of its two children with the hash of the transactions as the leaves of the tree.
Including the tree's root links the transactions to the PoW and the final block.
Now the miner's goal is to find a nonce such that the PoW output (a hash) meets the global difficulty (here, literally the product of the hash multiplied by the difficulty must be smaller than 2\textsuperscript{256}).
Thus, a miner needs to draw a new nonce and recompute the hash until it satisfies this goal.
The network can easily verify that the proof holds through a single round of hashing, and by including the block in the blockchain, rewards the miner with the block reward expressed through the Coinbase transaction.
When using mining pools, the pool pushes jobs (containing the PoW input) asking the miners participating in the pool to find a nonce that satisfies a lower difficulty than that of the total network.
When this lower difficulty is met, the miner is awarded a share of the final block reward and if by chance the actual difficulty is also met, the pool mined a block.

\section{Prevalence of Browser Mining}
\label{sec:world-wide}
We start our analysis of browser-based mining by investigating its prevalence in the web.
Thus, we visit landing pages of a large body of domains and identify the presence of mining code using two approaches.
Initially, we use a light-weight approach to download website landing pages via TLS across several datasets, \ie{} .com ($\sim$116M), .net ($\sim$12M), .org ($\sim$9M), and Alexa Top 1M ($\sim$950K), and match their HTML body against a public filter list (\sref{sec:nocoin}).
Subsequently, we instruct a Chrome browser to visit a subset of these domains to execute the webpage code and thereby monitor Websocket interactions and WebAssembly (Wasm) code as prevalent techniques for browser-based mining (\sref{sec:chrome}).
We obtain our datasets through DNS resolutions~\cite{netray} from zone files available at Verisign~\cite{verisignTLD} (.net/.com) and PIR~\cite{pir} (.org).

\subsection{NoCoin List}
\label{sec:nocoin}

\begin{figure}
\centering
\includegraphics{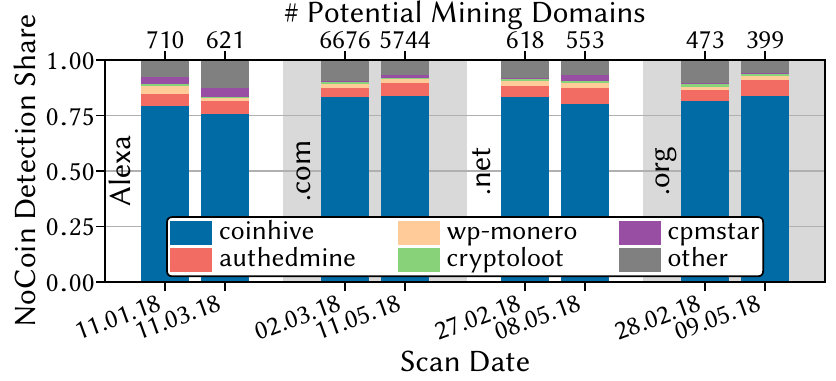}
\vspace{-2em}
\caption{NoCoin detected miners on the Alexa Top 1M and .com/.net/.org domains.}
\label{fig:nocoin}
\vspace{-1em}
\end{figure}

We visit every domain, prefixed with \texttt{www.}, via TLS and download the first \unit[256]{kB} of the domains' landing pages using \texttt{zgrab}.
\unit[256]{kB} offers a good tradeoff between capturing most content (\ie{} scripts are often located in the head of the document) and having a point where to stop downloading when pages do not stop sending data.
We then extract all javascript tags using \texttt{lxml} to apply the NoCoin filter list~\cite{nocoin}.
This list contains regular expressions to detect and subsequently block mining code using common ad blockers.
 \fig{fig:nocoin} shows the number of domains with hits to NoCoin filter rules on the top x-axis.
Relative to the number of domains, the bars on the y-axis show the relative share of the top 5 mining scripts (multiple per website possible).
We find the prevalence of mining websites to be rather low.
Yet in comparison, (popular) Alexa-listed domains have the largest share (up to 0.07\%).
This seems likely since mining is most profitable with websites having many visitors.
Looking at the miners, we find Coinhive, a Monero-based miner to be most prevalent (used by $>$\,75\% of the mining sites).
Notably, Authedmine, a variant of Coinhive asking for explicit user consent to mine and \emph{wp-monero} a WordPress plugin follows but at much lower shares.
We find other miners with smaller shares, \eg{} Cryptoloot a Coinhive clone.
By manually inspecting a random subset, we find false positives, \eg{} \emph{cpmstar} is a gaming ad-network that we could not verify to contain mining code.
For the once popular jsMiner~\cite{jsminer}, we find only 31 instances in all datasets combined.

\takeaway{We observe a low prevalence of mining in landing pages according to the NoCoin list. Most miners are Monero-based with Coinhive having the largest share ($>75\%$).}

\subsection{Chrome}
\label{sec:chrome}
We complement the NoCoin analysis by broadly investigating patterns of mining behavior when actually {\em executing} the pages.
This enables to find mining domains beyond NoCoin-listed pattern.
Through manual miner code inspection, we find that the majority of javascript miners utilizes WebAssembly (Wasm) for efficient PoW calculation.
WebAssembly~\cite{wasm} is a binary instruction format---featured in recent browsers---that enables to compile \eg{} C-code to Wasm for efficient execution within the browser. 
Further, the communication to the backend servers providing the PoW input often uses Websockets.
To detect these, we instrument a stock Chrome web browser using the Chrome Dev Protocol~\cite{chromedevprotocol} to capture all Websocket communication and to dump all detected Wasm code.
To decide when a page is fully loaded, we wait for the page's load event and set a \unit[2]{s} timer on every DOM change but wait no longer than additional 5\,s before we mark the page as loaded completely.
In case of no load event, we wait no longer than \unit[15]{s} to mark the website as timed out.
We further save the first \unit[65]{kB} of the final HTML to enable comparison with the NoCoin list used previously.

\begin{table}[t]
\centering
\renewcommand{\arraystretch}{.9}
\small
\begin{tabular}{@{}l|lr|lr@{}}
\toprule
    & \multicolumn{2}{c|}{Alexa} & \multicolumn{2}{c}{.org} \\[-0.5em] %
        & Class. & Count  &Class. & Count  \\ \midrule %
1 & coinhive         & 311     & coinhive          & 711  \\ %
2 & skencituer       & 123     & cryptoloot        & 183  \\ %
3 & cryptoloot       & 103     & web.stati.bid     & 120          \\ %
4 & UnknownWSS       & 56      & freecontent.date  & 108  \\ %
5 & notgiven688      & 46      & notgiven688       & 92 \\ \midrule %
Total & WebAssembly & 796 & WebAssembly & 1491  \\ \bottomrule
\end{tabular}
\vspace{0.5em}
\caption{Top 5 ($\sim$80\%) WebAssembly signatures. Most WebAssembly are miners ($\sim$96\%), dominated by Coinhive.}
\label{tab:wasm}
\vspace{-2em}
\end{table}

\afblock{Measurements.}
As this measurement is more time consuming, we restrict our scope to the .org zone and the Alexa 1M.
We prefix every domain with \texttt{http://www.}\ allowing Chrome to follow redirects to the secured variant if necessary.
Thus in contrast to our previous TLS-only measurement, we also analyze non-HTTPS websites.
We build signatures from the Wasm code by combining (in a strict order) and then hashing the contained functions with SHA256.
Through manual inspection of the Wasm, we build up a database of $\sim$160 different assemblies (often versions of the conceptually same Miner) that we found and categorized them, \eg{} through their Websocket communication backend or by some other distinguishing feature that we found in the code.
Such features \eg{} comprises the number of XOR, shift or load operations which we found to be quite distinctive or function name hinting at the hash function itself.

\begin{table}[t]
\centering
\setlength\tabcolsep{2pt}%
\renewcommand{\arraystretch}{.9}
\small
\begin{tabular}{@{}c|cc||ccc@{}}
\toprule
      & \begin{tabular}[c]{@{}c@{}}  \small NoCoin\\ \small Hits \end{tabular} & \begin{tabular}[c]{@{}c@{}} \small having Wasm\\ \small Miner \end{tabular}& \begin{tabular}[c]{@{}c@{}}  \small Wasm\\ \small Hits \end{tabular} & \begin{tabular}[c]{@{}c@{}} \small blocked\\ \small by NoCoin \end{tabular} & \begin{tabular}[c]{@{}c@{}} \small missed\\ \small by NoCoin \end{tabular} \\
      \midrule
Alexa & 993        & 129                      & 737            & 129  &  608 (82\%)                     \\
.org  & 978        & 450                      & 1372           & 450   & 922  (67\%)            \\ \bottomrule    
\end{tabular}
\vspace{0.5em}
\caption{Miners on Chrome data (incl.\ non-TLS) found through NoCoin and by our WebAssembly signatures.}
\label{tab:nocoinvswasmsig}
\vspace{-1em}
\end{table}

Table~\ref{tab:wasm} summarizes our findings for the Alexa 1M and the .org TLD from measurements in the first two weeks of May 2018.
We observe most Wasm code to contain mining functionality and most miners are again Coinhive.
To put the Chrome-based approach in perspective to the NoCoin list, we apply the NoCoin block list to HTML saved by Chrome, \ie{} after the execution of javascripts.
Table~\ref{tab:nocoinvswasmsig} shows the number of miners detected by the NoCoin list and the fraction of mining Wasm on this part as well as the total number of websites classified through our Miner Wasm signature database and the subset of websites that were detected by the NoCoin list.
We observe that NoCoin classifies many websites as miners, of which only a fraction actually embeds mining Wasm code.
This indicates false positives which we verified through random inspections.
If we take a look at the websites for which we found Wasm mining signatures, again, the NoCoin list only classifies a fraction of these as having a miner---false negatives.

\begin{table}[t]
\centering
\setlength\tabcolsep{2.2pt}%
\renewcommand{\arraystretch}{.9}
\small
\begin{tabular}{@{}l|lrlr||lrlr@{}}
\toprule
      & \multicolumn{4}{c||}{Alexa}                             & \multicolumn{4}{c}{.org}                               \\[-0.5em]
      & \multicolumn{2}{c}{NoCoin} & \multicolumn{2}{c||}{Signature} & \multicolumn{2}{c}{NoCoin} & \multicolumn{2}{c}{Signature} \\
      \midrule
1     & Gaming                     & 19\%         & Pornogr.                    & 19\%        &    Gaming           &      29\%       &      Religion       &    9\%       \\
2     & Edu. Site                    &   9\%         &  Tech.                    &   8\%         &    Business         &      8\%        &      Business       &    8\%        \\
3     & Shopping                    &   8\%        &   Filesharing          &   8\%         &     Edu. Site        &      6\%         &     Edu. Site         &  8\%          \\
4     & Pornogr.                        &    7\%        &   Edu. Site             &   5\%         &    Pornogr.             &      5\%        &    Health Site        &   7\%         \\
5     & Tech.                          &    6\%        &   Ent. \&  Music   &   5\%          &    Shopping        &     4\%         &   Tech.                 &   6\%         \\
\midrule
\multicolumn{2}{l}{Categorized}  & \multicolumn{1}{c}{79\%}       & \multicolumn{2}{c||}{74\%}     & \multicolumn{2}{c}{54\%}         & \multicolumn{2}{c}{42\%}     \\ \bottomrule
\end{tabular}
\vspace{0.5em}
\caption{Top 5 categories according to Symantec RuleSpace.}
\label{tab:miner_cats}
\vspace{-2em}
\end{table}

\afblock{Classification.}
We use the Symantec RuleSpace\footnote{Used in Symantec products to classify websites.}~\cite{rulespace} engine to categorize the mining websites.
Table~\ref{tab:miner_cats} shows the top 5 categories to which RuleSpace assigned the websites for the NoCoin list matches as well as our signature-based approach.
We observe a diverse set of categories and that RuleSpace can classify a larger body of Alexa domains than .org domains.
Interestingly, the categories for NoCoin and our approach differ, especially the top category shows a large mismatch, \ie{} Gaming vs.\ Pornography and Gaming vs.\ Religion.
This could be caused by the aforementioned gaming ad-network.

\takeaway{Miners are already embedded on websites today. 
Simple block lists are ineffective to detect them all and our signature-based approach can detect sites beyond the NoCoin block list.
Still, Coinhive is the most used mining service.}

\section{The Coinhive Service}
\label{sec:coinhive}
Coinhive provides a mining service advertised with the slogan \emph{``Monetize Your Business With Your Users' CPU Power''}~\cite{coinhive}, we observed Coinhive to have the most widespread use (see Section~\ref{sec:world-wide}).
Their services are built on providing a highly optimized Monero javascript miner to be embedded in websites.
In turn, Coinhive keeps 30\% of the mined reward.
Apart from offering this API, Coinhive offers \eg{} a Captcha service and a short link forwarding service which is the subject of our first analysis.
Our tools on which the following analysis is based are available at~\cite{coinhiveDataset}.

Regardless of the actual service, the process works as follows:
\emph{i)} A Coinhive user (\eg{} a website owner) is assigned a unique token that is included in the API calls which is used to associate the mined shares.
\emph{ii)} Upon a website visit, the miner is loaded and connects to the Coinhive pool and authorizes with the user's token to receive input for hashing.
\emph{iii)} Once a valid hash is found, it is committed to the Coinhive pool.
\emph{iv)} Eventually, Coinhive pays their users 70\% of the block reward and keeps the remaining 30\%.

\subsection{Short Link Forwarding Service}
\label{sec:fwd}
To begin analyzing Coinhive, we focus on its short link forwarding service, which is similar to a common short link service (\eg{} bit.ly) but additionally requires to compute a configurable number of hashes before resolving the link.
When a user visits a link, she only sees a progress bar indicating the share of hashes that have been computed, when all locally computed hashes have been sent to the service (\ie{} the progress bar is full), the service will return the original link and will instruct the browser to redirect the user to it.
This link redirection monetization is comparable to short link services delaying the redirection while serving advertisements and paying the link creator a commission~\cite{stranger_danger}.
With Coinhive, the creator of the short link receives a share of the block reward that is mined by the users visiting the short links.

Their short links follow a simple structure, identified by an alphanumeric ID: \texttt{https://cnhv.co/[a-z0-9]}.  %
We observed that new links are assigned increasing IDs which enables one to enumerate the link address space.
As of February 2018, up to 4 characters are used, resulting in a total of 1,709,203 active short links.
We visit all links and gather the Coinhive redirection HTML document to collect {\em i)} the link creator's token---used to associate performed hashes to the link creator---as well as {\em ii)} the number of hash computations required by the link creator to resolve the link.
Even though a single user could own multiple tokens, we will regard users and tokens as synonymous in this paper.

\begin{figure}
\centering
\includegraphics[]{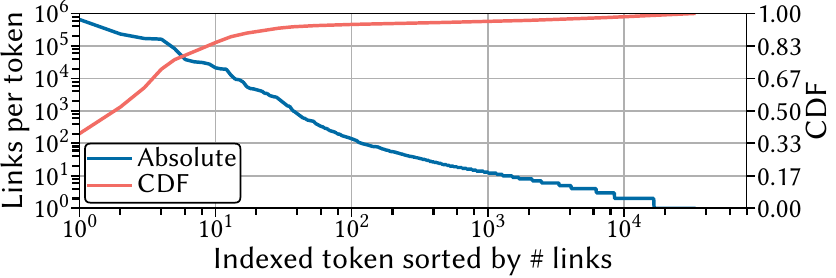}
\vspace{-1.5em}
\caption{The number of links per token (users) is heavily biased towards a small number of users.}
\label{fig:tokens_distribution}
\end{figure}

Without actually computing hashes, we can already look at {\em i)} the distribution of short links per Coinhive users as well as {\em ii)} the required number of hashes to resolve the links.
\fig{fig:tokens_distribution} shows the distribution of short links per user.
We observe a power-law which highlights the existence of a few heavy users that created a large number of links.
In fact, 1/3 of all links are contributed by a single user only and roughly 85\% of all links are created by only 10 users.
Of course, a single user could use multiple tokens, however, this would only emphasize our current observations.

\begin{figure}
\centering
\includegraphics[]{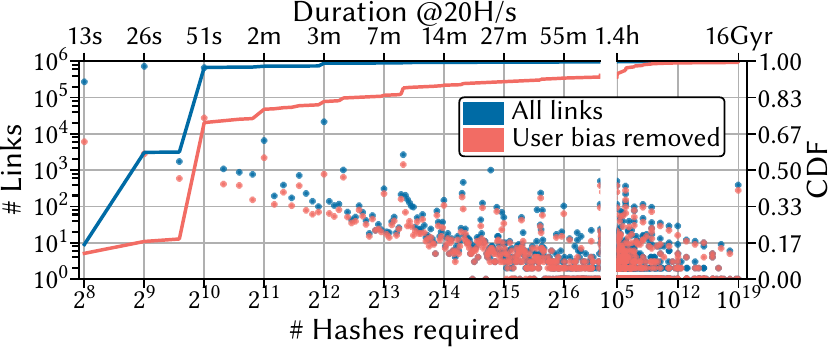}
\vspace{-1.5em}
\caption{Required number of hashes and their frequency of occurrence as well as the time it takes to compute these hashes. Please note the skewed x-axis.}
\label{fig:targets}
\end{figure}

To actually resolve the link, the user needs to compute the required number of hashes set by the link creator.
\fig{fig:targets} shows the distribution of this link resolution difficulty in the number of required hash computations.
The blue (dark) portion of the CDF depicts all observed links, while the red (light) portion removes the previously observed bias by heavy user by counting a required \#\,hashes only once per user; \ie{} 1000 links from the same user with the same number of required hash computations are only counted once instead of 1000 times as in the blue (dark) dataset.
To provide a perspective on the time it takes to resolve a short link, the top x-axis shows the duration to compute the required \# of Cryptonight hashes in a Chrome browser with a commodity laptop\footnote{2013 Macbook Pro \unit[2.8]{GHz} Intel Core i7: \unit[20]{h/s} with 4 threads.}.
We observe that the majority of links can be resolved in less than 51\,sec (1024 hashes).
The heavy user bias is most prominent at 512 hashes, still, when removing the user-bias over 2/3 of the links of all users can be solved with at most 1024 hashes in below one minute.
To our surprise, many links require a longer time to resolve; we find many different users and over hundreds of short links that set the maximum of 10\textsuperscript{19} hashes which takes several billion years to resolve.
While the user's willingness to wait certainly depends on the content that is supposed to be behind a short link, high values suggest either no desire to have them ever resolved or misconfigurations (\eg{} short link creators are not aware of the actual duration).

\afblock{Link Destinations.}
To understand the kinds of links that the short link service is used for, we resolve all links which require less than 10K hashes from the unbiased dataset (covering 85\% of this dataset see red (light) CDF in \fig{fig:targets}).
Additionally, we resolve a random sample of 1000 links for each of the top ten Coinhive users.
To efficiently resolve the short links without a web browser, we replicate the working principle of the web miner in a non-web implementation that can resolve multiple short links in parallel making use of the official optimized Monero hash code.
We found that Coinhive alters the block header contained in the PoW inputs before sending them to the users which the web miner reverts deep within its WebAssembly\footnote{A simple XOR with a fixed value at a fixed offset}.
This appears to be a countermeasure to prevent using the Coinhive web miner outside of the Coinhive environment, e.g., in custom mining pools.
We had to roughly compute 61.5M hashes which we were able to do in little less than two days on a capable server machine.

\begin{table}[]
\centering
\setlength\tabcolsep{1.8pt}
\small
\renewcommand{\arraystretch}{.9}
\begin{tabular}{@{}lll||lll@{}}
\toprule
Domain & Category   & Freq. & Domain & Category &   Freq. \\ \midrule
youtu.be  & Ent. \& Music & 20\%  & ftbucket.info & Msg. Board &  9.9\%\\ 
zippyshare.com & Filesharing & 10\% & getcoinfree.com  & Finance & 9.2\%\\ 
icerbox.com & Filesharing & 10\% &  ul.to  & Filesharing & 4.2\% \\ 
hq-mirror.de & Ent. \& Music & 10\% & share-online.biz & Filesharing &  2.9\% \\ 
\begin{tabular}[l]{@{}l@{}}andyspeed\\racing.com\end{tabular}  & Automotive & 10\% &  \begin{tabular}[l]{@{}l@{}}oboom.com\\  \phantom{x} \end{tabular}& \begin{tabular}[l]{@{}l@{}} Filesharing \\  \phantom{x} \end{tabular}&   \begin{tabular}[l]{@{}l@{}} 2.8\% \\ \phantom{x} \end{tabular}\\  \bottomrule
\end{tabular}
\vspace{1em}
\caption{Top 10 domains in 89\% of all samples from the top 10 short link creators.}
\label{tab:top10}
\vspace{-2.4em}
\end{table}

\afblock{Top 10 User.}
We first investigate a random sample from all short links of the top 10 users (1000 links each) representing 80\% of the link targets.
Table~\ref{tab:top10} shows a classification for the top 10 domains (accounting for roughly 89\% of all sampled URLs) that we extracted from the destination URL.
We again utilize the RuleSpace categories to manually classify those 10 domains.
As the table shows, most links point to streaming and filesharing services.

\afblock{Top Categories.}
We employ the RuleSpace engine to further classify the unbiased dataset into categories.
One URL can have multiple categories, therefore, a single URL can contribute to different categories.
For roughly 1/3 of the URLs RuleSpace has no classification, for the remainder, Table~\ref{tab:top_cats} lists the top 10 categories and how often a URL falls into each category.
We observe that sites to fall into a diverse set of categories, unlike the top 10 users for which filesharing and streaming were the dominant categories (Table~\ref{tab:top10}). 

\takeaway{Coinhive's link forwarding service is dominated by links from only 10 users. 
They mostly redirect to streaming videos and filesharing sites.
We find that most short links can be resolved within minutes, however, some links require millions of hashes to be computed which is infeasible.}

\subsection{Estimating the Network Size}
While we find many websites to use Coinhive (see \sref{sec:world-wide}), it remains unclear how many users visit these sites.
Thus, the mining power and the achievable payouts are unknown.
To understand the available mining power and thereby the users of Coinhive, we need to identify which blocks in the Monero blockchain were mined through Coinhive.

\afblock{Methodology.}
When a block is mined by the Coinhive network, one of the clients must have found a nonce that satisfies the PoW difficulty.
Then, a new block can be mounted into the blockchain which contains the block header that is also part of the PoW input, as well as all the transactions that have implicitly been included in the PoW input through the Merkle tree root (see \fig{fig:monero_overview}).
Thus, if we find the PoW input for which a suitable nonce was found, we can investigate the blockchain and look at the block that succeeds the block referenced in the PoW.
If the transactions in that block form a Merkle tree whose root is equal to that in the PoW input, we can be sure that the PoW input was the one that was used to mine the block.
This uniquely identifies the origin as each block contains the Coinbase transaction (first leaf of the Merkle tree) which is used to pay the block rewards to the miner (\ie{} Coinhive).
Thus we could never by accident see a Merkle tree root of another miner in the PoW input.

\begin{table}[]
\centering
\renewcommand{\arraystretch}{.9}
\small
\begin{tabular}{@{}ll||ll@{}}
\toprule
Category   & Count & Category &   Count \\ \midrule
\CategoryOneName{}    & \CategoryOneOcc{}         & \CategorySixName{}        & \CategorySixOcc{}                             \\
\CategoryTwoName{}    & \CategoryTwoOcc{}          & \CategorySevenName{}   & \CategorySevenOcc{}                             \\
\CategoryThreeName{} & \CategoryThreeOcc{}        & \CategoryEightName{}     & \CategoryEightOcc{}                             \\
\CategoryFourName{}   & \CategoryFourOcc{}          & \CategoryNineName{}      & \CategoryNineOcc{}                             \\
\CategoryFiveName{}   & \CategoryFiveOcc{}           & \CategoryTenName{}        & \CategoryTenOcc{}                             \\ \bottomrule
\end{tabular}
\vspace{1em}
\caption{Top 10 categories of the unbiased dataset $<$\,10K hashes.}
\label{tab:top_cats}
\vspace{-2em}
\end{table}

We investigate the PoW inputs that are delegated by Coinhive to its users by connecting to one of their mining pools and request a new PoW input every \unit[500]{ms}.
As the network finds a new block on average every two minutes, we cluster the PoW inputs by the pointer to the previous (at time of reception, most recent) block.
We found that we never obtain more than 8 different PoW inputs (even though more exist theoretically).
Coinhive currently operates 32 mining endpoints (which can be gathered from the javascript or by enumerating the domain name), when we connect to all of them and repeat the process, we observe at most 128 different PoW inputs per block.
While this suggests that there are two endpoints per backend system, it also puts us into the position to actually investigate each of the 128 PoW inputs and check the Merkle tree root against the Merkle tree root of the transactions in the block that was actually mined after that referenced in the PoW input.

\begin{figure}
\centering
\includegraphics[]{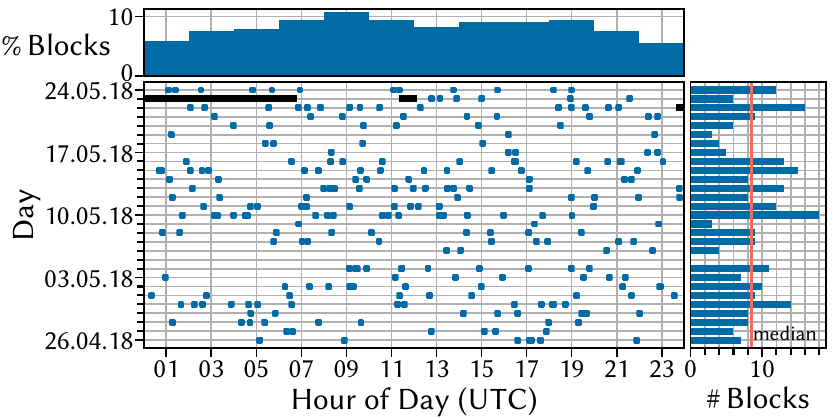}
\vspace{-2em}
\caption{Mined blocks over time from the Coinhive network. Black parts mark outages of \emph{our} infrastructure.}
\label{fig:mined_blocks}
\vspace{-1em}
\end{figure}

\afblock{Measurements.}
We have been requesting new PoW inputs for four weeks and we are thus able to confidently estimate a lower bound on the blocks mined through Coinhive.
\fig{fig:mined_blocks} shows a blue block for every Coinhive-mined block as well as the total number of blocks on that day.
As finding blocks correlate with users mining through Coinhive, we were interested to see if blocks are found at certain times which could hint at the geolocation of the users.
However, the figure (upper subplot) shows that blocks are found throughout the whole day which might be an indicator of the global reach of Coinhive.
We find multiple days with significantly more blocks than on average, \eg{} the 30th of April, 10th and 22nd of May 2018.
The 30th of April precedes Labor Day, a public holiday in over 80 countries, time zone shifts to UTC or holidays could explain increased Internet usage resulting in more mined blocks.
Similarly, the latter two were Ascension Day and the day after Pentecost, both public holiday in many (mostly) European countries.

In the median (average), we find \CoinhiveMedBlocksPerDay{} (\CoinhiveAvgBlocksPerDay{}) blocks per day, we noticed a disruption of Coinhive' service on the 6th and 7th of May which resulted in only a few to no announced PoW inputs.
We can estimate the combined hash rate of Coinhive by taking the network's difficulty into account.
The difficulty denotes the expected number of hashes that are required to find a block which is adjusted after each block such that the block rate of two minutes is met.
Over the course of our observations, the median difficulty was \MoneroMedDiff{} hashes, which translates to a network hash rate of \unit[\MoneroMedHashRate{}]{h/s}.
As Coinhive mines roughly \CoinhiveMedBlocksPerDay{} blocks per day, they produce \CoinhiveMedBlocksPerDayShare{} of all 720 blocks/day which translates to \unit[\CoinhiveMedHashRate{}]{h/s}.
If we assume that a web client performs between 20 to 100 h/s, then Coinhive requires between \CoinhiveAvgUsersLowPerf{} and \CoinhiveAvgUsersHighPerf{} constantly mining users.
Compare our findings with numbers reported by Coinhive~\cite{coinhiveFirstWeekBlog} from September 2017 is difficult. 
Coinhive wrote that their hash rate peaked at \unit[13.5M]{h/s} (then 5\% of the network's hash rate).
However, our results are averages over long periods of time and derived from the statistical properties of the network, while those published are momentary peak rates, thus a direct comparison is not possible.

\begin{table}[]
\renewcommand{\arraystretch}{.9}
\begin{tabular}{c|cc|c|c}
\toprule
     & Med. & Avg.  &  Hashrate & Currency \\
     & \multicolumn{2}{c|}{[blocks/day]} & [MH/s] & [XMR] \\
     \midrule
May  &        \medblocksperday{may} &\avgblocksperday{may}         &   \medianhashrate{may}   &  \xmrminded{may}          \\
June &         \medblocksperday{june}& \avgblocksperday{june}            &     \medianhashrate{june} &  \xmrminded{june}                 \\
July &        \medblocksperday{july} &\avgblocksperday{july}            &     \medianhashrate{july} &   \xmrminded{july}              \\ \bottomrule
\end{tabular}
\caption{Coinhive mining statistics for three month in 2018.}
\label{sec:monthstats}
\vspace{-3em}
\end{table}

If we sum up the block rewards of the actually mined blocks over the observation period of four weeks, we find that Coinhive earned \unit[\CoinhiveXMRRewardPerMonth{}]{XMR}.
Table~\ref{sec:monthstats} complements our four-week analysis with three full months in 2018 showing its continuity.
Similar to other cryptocurrencies, Monero's exchange-rate fluctuates heavily, at the time of writing one XMR is worth 120 USD, having peaked at 400 USD at the beginning of 2018.
Thus, assuming 120 USD, Coinhive mines Moneros worth around 150,000 USD per month of which they say, they give 70\% to their users.
Still, the operational costs seem manageable, making it potentially profitable for Coinhive.

\takeaway{
Coinhive currently contributes $\sim$\CoinhiveMedBlocksPerDayShare{} of the mining power of the Monero network.
While probably profitable for Coinhive, it remains questionable whether mining is a feasible ad alternative.}

\section{Related Work}
\label{sec:rw}
Browser-based mining has been subject to substantial media coverage, \eg{} reports on Pirate Bay~\cite{guardianPirateBay} mining, about hacked websites for mining~\cite{bbcMiningHacking}, miners injected into the Google's DoubleClick ad-platform~\cite{trendMicroGoogleAds} or drive-by Monero mining on Android~\cite{malewarebytesBlogAndroid}.
Many blog posts exist that report on Alexa-listed websites to include mining code~\cite{pixalateAlexaBlog, adGuardAlexaBlog}, however, without detailing a methodology.
A list published on Github~\cite{githubCoinhiveBadList} provides an overview of potential mining domains.
However, this list also includes entries such as \textit{google.com} which is unlikely to be mining.
To the best of our knowledge, \cite{eskandari2018} are the first to academically investigate browser-based mining parallel to our work.
While they also find Coinhive to be the most prominent service, their analysis is based on a string search over a large set of HTTP bodies, thus not accounting for actual HTML structure which we did in our first analysis (see \sref{sec:nocoin}) which showed to already produce skewed results, \eg{} the categories of websites significantly differs.
We thus complement their results by incorporating WebAssembly fingerprinting and further shed light on the inner workings of Coinhive. 
In concurrent work, Konoth et al.~\cite{minesweeper} estimate revenue for websites on the Alexa Top 1M including mining scripts.
Further, they also analyze the nature of Wasm mining for detection using a very similar methodology to what was the basis for our fingerprinting such as counting certain instruction or memory access pattern.
Based on monitoring DNS, \cite{360labsDNS} also observes Coinhive as the dominant player.
They report that most crypto miners are present on adult websites in the Alexa Top 1K/3K.
Similar with regards to link-forwarders, \cite{stranger_danger} analyzed ad-based link forwarding services and their revenue model which relates to that of Coinhive, thus we believe their results to also apply here.

\section{Conclusion}
\label{sec:conclusion}
This paper analyzes the prevalence of browser-based mining, a new revenue generating model to monetize websites and an alternative to ad-based financing that is enabled by ASIC resistant cryptocurrencies such as Monero.
By inspecting 137M .com/.net/.org and Alexa Top 1M domains for mining code, we indeed find websites that utilize browser-mining.
Yet, the prevalence of browser mining is {\em currently} low at $<0.08\%$ of the probed sites. 
For its detection, we find the public NoCoin filter list to be insufficient to broadly detect browser mining.
We thus present a new technique based on WebAssembly fingerprinting to identify miners, up to 82\% of thereby identified mining websites are not detected by block lists.
We identify Coinhive as the largest web-based mining provider used by 75\% of the mining sites.
Given its popularity, we further dissect Coinhives' link-forwarding service.
We find that 10 heavy users contribute over 80\% of all short links mostly targeting streaming and filesharing services.
The remaining short links target a diverse set of websites.
We continue by dissecting the economics of Coinhive, we devise a new method that allows associating mined blocks with a mining pool and we find that Coinhive mines \CoinhiveMedBlocksPerDayShare{} of all Monero blocks and their visitors have a combined median hash rate of \unit[\CoinhiveMedHashRate{}]{h/s}.
While we find that Coinhive turns around Moneros worth 150,000 USD per month, the current value stability of cryptocurrencies requires further investigations if browser-based mining can be an alternative revenue model to ad-based financing.
Further, the impact of the CPU intensive miner on a website's performance, a mobile device's battery lifetime or a visitor's energy bill is yet to be quantified but it could be a huge hurdle to be competitive to ad-based financing on a larger scale.

\begin{acks}
Funded by the Excellence Initiative of the German federal and state governments, as well as by the German Research Foundation (DFG) as part of project B1 within the Collaborative Research Center (CRC) 1053 -- MAKI.
Further, we thank Martin Coughlan from Symantec for granting us access to the RuleSpace classification engine as well as the network operators at RWTH Aachen University, especially Jens Hektor and Bernd Kohler.
\end{acks}

\bibliographystyle{ACM-Reference-Format}
\balance
\bibliography{literature} 

\end{document}